# An analytical study of content and context of keywords on physics

Bidyarthi Dutta

Assistant Professor, Department of Library and Information Science, Vidyasagar University Midnapore 721 102, West Bengal,
Email: bidyarthi.bhaswati@gmail.com

This paper is based on the analysis of author-assigned and title keywords and their constituent component words collected from 769 articles published in the journal *Low Temperature Physics* since the year 2006 to 2010. The total number of distinct keywords is 1155 of which 869 are single keywords having total frequency of occurrence of 2287. The single keywords have been categorized in four broad classes, viz. eponymous word, form word, acronym and semantic word. A semantic word bears several contexts and thus it may be considered as relevant in several other subject areas. The probable subject areas have been found with the aid of two popular online reference tools. The semantic words are further categorized in twelve classes according to their contexts. Some parameters have been defined on the basis of associations among the words and formation of keywords in consequence, i.e. Word Association Density, Word Association Coefficient and Keyword Formation Density. The values of these parameters have been observed for different word categories. The statistics of word association tending keyword formation would be known from this study. The allied subject domains also become predictable from this study.

## Introduction

The existing knowledge organization systems, by and large solicit controlled indexing language or controlled vocabularies. But its suitability regarding optimum recall and precision values has been debated over decades. The uncontrolled vocabulary systems, though very popular today, still not full-proof regarding the question of standardization. This paper studies the uncontrolled vocabulary system of a specific subject 'Low temperature physics' through the assigned keywords of the research articles. A keyword may belong to multiple subject domains and may be formed by one or more number of words, calling precisely single-worded keyword and multi-worded keyword respectively. Now if a multi-worded keyword belongs to a particular subject domain(s), then its constituent words may or may not belong to that particular subject domain. This aspect is addressed in this paper. The contexts in terms of subject-domains of the words in keywords belonging to *low temperature physics* are studied.

## Review of literature

Svenonius[1] discussed that controlled vocabularies "bring like things together" to facilitate access and discoverability. The critics of traditional systems accuse controlled vocabularies for being artificial and representing a biased view of the structure of the universe of knowledge[2,3]. Svenonius[4], Fidel[5] and Rowley[6] put several arguments on inappropriateness of controlled vocabulary system. According to Noruzi[7], uncontrolled terms or tags or keywords are words or phrases users attach to resources that may help in later retrieval. Lu[8] presents several advantages to the use of uncontrolled terms. White[9] carried out a comparative study between controlled vocabularies and free text keywords. Engelson[10] studied correlations between title keywords and LCSH terms. The usefulness of keywords in science journals was described by Hartley[11]. The structured keyword method for increasing information sharing among scientists was proposed by Kajikawa[12]. Gil-Leiva[13] examined author keywords from scientific articles and found a 46% overlap with subject headings when author keywords were normalized. Frost[14] studied the correlation between LCSH terms and derived keywords from titles in bibliographic records. Ansari[15] carried out a comparative study between assigned descriptors and title keywords in medical theses. Voorbiz[16] carried out comparative study between title keywords and subject descriptors in humanities and social sciences. Strader[17] executed comparative study between author keywords and Library of Congress subject headings. Gross[18] analyzed the effect of controlled vocabulary on keyword searching. Kipp[19,20] examined author keywords in comparison to tags and subject headings using a modification of Voorbij's categories[16] and found a high degree of overlap between tags, author keywords and subject headings. Kipp[21] observed tagging practices on CiteULike. Schultz[22] compared author keywords to document titles and to indexing terms assigned by subject matter experts and found author keywords matched subject terms more closely than title terms.

Heckner[23] studied tags and author keywords and found an approximately 58% overlap in content. Montgomery[24] observed high degree of concurrence between title keywords for entries in Index Medicus and assigned subject headings (86%), but found 14% of articles were not indexable based solely on the title. Carlyle[25] compared user vocabulary directly to LCSH and found 47% exact match between user vocabulary and LCSH and up to a 70% match when using stemming and other matching algorithms to correct for plurals and punctuation. O'Connor[26] found that many indexes had much lower rates of match between title keywords and subject headings. Garrett[27] studied the use of subject headings to enhance eighteenth century documents and found that as many as 60% of searches would fail without the addition of keywords due to terminological drift over time. Davarpanah[28] examined the relative effectiveness of title keywords and assigned subject descriptors in representing the content of theses in the Iranian Dissertations Database. Huang[29] discussed how syllable or word division in bibliographic records of Chinese materials affects title keyword searches. Jahoda[30] tested searching of 3204 documents in the field of chemistry by selecting keywords from title index and alphabetic subject index. Adams[31] executed a comparative study between keywords in title and cited references. Diener[32] measured the informational value of journal article titles by counting number of words in titles and title keywords. Alvarez[33] designed a method for measuring information from keywords, using the Rasch model as the measuring instrument.

Hurt[34] examined the differences between author-keywords and automatically generated keywords for polymer science literature. Gbur[35] framed suitable guidelines for the selection of optimal keywords in the subject field of statistics. Tillotson[36] raised question on utility of keywords as searching tag. Wellisch[37] remarked the significance of *keyword* only as subject descriptor. Craven[38,39,40] studied variations in use of meta-tag keywords and meta-tag descriptors by web pages in different subjects and different languages. Craven[41] also discussed role of keywords in meta-tagging of web page descriptions. Turney[42] developed algorithms for automatic selection of important, topical phrases or keyphrases from within the body of a document. Kishida[43] developed statistical methods for automatically assigning classification numbers and descriptors based on title keywords of journal articles. Jones[44] studied automatic keyphrase extraction methods for use in digital libraries. Taghva[45] explored the use of manually assigned keywords for query expansion with interactive tools. Automatic keyword extraction methods in specific subject domains were explored by Frank[46]. Hulth[47] discussed automatic keyword or keyphrase extraction process from linguistic point of view. Cleverdon[48] showed that each indexing system was made up of a basic vocabulary system.

The literature review shows no research done till date that involves dismantling a keyword into its constituent root words for analysis. The keywords are usually made up of one or more root word(s) with word stem (optional), combined with modifiers. It is necessary to analyze the context and semantic features of the constituent root words to understand the multi-contextual features (if any) of the keyword. Also, it needs study whether the contextual and semantic features of the constituent root words differ from the same for the keyword, which is not yet investigated and forms a research gap in keyword research. This study tries to bridge the gap in keyword research and presents an analytical model for study of subject-specific keywords.

**Objectives of the study**

- To dislodge author-assigned and title keywords consisting of more than one words into constituent components to collate all single words and keywords together for studying the statistics of occurrence of the same;
- To categorize analyzed words in accordance with the modes of occurrences;
- To find out the values of three fundamental variables, i.e. frequency of words (f), number of associations among words (a) and number of keywords formed (k), which are linked with the modes of occurrences of words and types of associations among the words;

- To define five parameters on the basis of these three fundamental variables in order to study the nature of association among the words while forming keywords and to find out numerical values of them; and
- To find out the subject areas for all semantic words with the aid of online dictionary and Wikipedia.

## Methodology

In this study, the author-assigned and title keywords are collected from 769 articles published in the journal entitled *Low Temperature Physics* between 2006 to 2010. The collected keywords consisting of more than one words were analyzed in constituent words. The subject areas of the constituent words were then found out with the aid of *Online Dictionary* and *Wikipedia*. The context of the constituent words of a keyword were obtained in this way. This study has been carried out by keywords selected from nine to thirteen years old articles and in the next phase, similar studies will be carried out by keywords of recent articles to observe whether any difference results. The number of author-assigned and title keywords collected from all these 769 articles are 1155 that constituted the sample for the study. The total frequency of these 1155 keywords is observed as 2280. These 1155 number of keywords have been analyzed into 869 numbers of single words having total frequency of occurrence as 2287.

The single words obtained from keywords have been categorized in four broad classes, viz. eponymous word, form word, acronym and semantic word. The words represent names of persons (proper noun) are categorized as eponymous words and represented by EW. The articles (a, an & the), prepositions and conjunctions are categorized as form words and represented by FW[49]. The abbreviations formed are categorized as acronyms and represented by AC and the remaining words that indicating any subject are categorized as semantic words and represented by SW. It is found that the semantic words are relevant to more than one subject areas. The relevant subject areas of each and every semantic word are found out with the aid of online dictionary[50] and Wikipedia[51]. These two online reference tools generally provide subject in context of any word, which is the reason for selecting them as an aid for this study. The number of relevant subject areas corresponding to each word is termed as degree of contextuality (D(C)) (Table 1). If no relevant subject area is found for a word in these two online reference tools, the same is termed as no-contextual word and represented by 0-C; similarly for only one relevant subject area this is mono-contextual word and represented by 1-C and so on.

For instance, the word relaxation is used in the context of the following three subject areas, viz. physiology, physics and mathematics as found in online dictionary. Similarly, according to Wikipedia, the relevant subject areas for the same word are physics, NMR, mathematics and psychology. A comparison between these two reference tools uniquely identifies the following subjects where the semantic word *relaxation* is considered as relevant, i.e. physiology, physics, mathematics, NMR and psychology. The keyword consisting of the word relaxation is shown in *italics* in Table 2. The word relaxation may be considered as a penta-contextual semantic word with the degree of contextuality five that may be represented as 5-C (Table 3).

Table 1 — Categories of semantic words (SW) by degree of contextuality (D(C))

| Word categories | No. of subject domains in context | Represented by |
|---|---|---|
| No contextual | 0 | 0-C |
| Mono-contextual | 1 | 1-C |
| Di-contextual | 2 | 2-C |
| Tri-contextual | 3 | 3-C |
| Tetra-contextual | 4 | 4-C |
| Penta-contextual | 5 | 5-C |
| Hexa-contextual | 6 | 6-C |
| Hepta-contextual | 7 | 7-C |
| Octa-contextual | 8 | 8-C |
| Nona-contextual | 9 | 9-C |
| Deca-contextual | 10 | 10-C |
| Higher-contextual | >10 | >10-C |

Table 2 — Occurrence of keywords over the years

| Keywords | Types of keywords (excluding form words) | Frequencies over years | | | | | Total |
|---|---|---|---|---|---|---|---|
| | | '06 | '07 | '08 | '09 | '10 | |
| Wide band gap semiconductor | Four worded keyword | 2 | | | 5 | 1 | 8 |
| *Nuclear spin-lattice relaxation effect* | *Five worded keyword* | | *1* | | | | *1* |
| Defect of absorption-spectra | Three worded keyword | 1 | | | | | 1 |
| Surface of acoustic wave | Three worded keyword | 1 | | 2 | | | 3 |
| Aharonov-Bohm effect | Three worded keyword | 1 | | | | 1 | 2 |

**Words in keywords: an analysis**

Suppose the following five keywords have been collected, viz. wide band-gap semiconductor, nuclear spin-lattice-relaxation effect, defects of absorption spectra, surface of acoustic-wave and Aharonov-Bohm effect. The occurrence statistics of these keywords over five years (2006-2010) is given in Table 2. For instance, the keyword Wide band-gap semiconductor appeared twice only in two articles out of 144 articles (Table 4) published in the year 2006, five times in five articles out of 149 articles published in 2009 and once in one article out of 155 articles published in 2010.

The total frequency of this keyword over the five years is thus eight. The frequency of a keyword in a particular year says the number of articles where the same appeared as any keyword has taken only once from an article. The numbers of words in this keyword is four, viz. wide, band, gap and semiconductor. The degrees of contextualities of these four words are 4, 5, 10 and 3 respectively as observed in online dictionary and Wikipedia. The degrees of contextualities of other words in keywords of Table 2 are presented in Table 3 along with number of keywords formed by each word. For instance, the word absorption appeared only in one keyword *Defect of absorption spectra*, whereas the word *effect* appeared in two keywords, viz. *Nuclear spin-lattice-relaxation effect* and *Aharonov-Bohm effect*. The types of each word are also indicated here. Only two words in Table 3, i.e. Aharonov and Bohm are eponymous words, as these words indicate the names of two physicists in the concerned subject domain. A physical fact or phenomenon is represented by one or more scientists' names, which is very common feature frequently observed in physics. The word *Of* in the Table 3 is a form word and all other words are semantic words with different degrees of contextualities (given in adjacent bracket) as they convey some sorts of meanings in relevant context.

**Wordship pattern: statistics of words in keywords**

The analytical study of the constituent words in keywords is carried out here. The name given to this study is wordship pattern, just in analogy with authorship pattern study in bibliometrics. The number of articles and keywords in different years are presented in Table 4. The number of distinct keywords for each year ranges roughly between 500 and 550 with an average number per article nearly 3.5 (Table 4). The overall average number of keywords per article is 1.5. The overall average is much less than yearwise average as a substantive number of keywords was repeated over years. The average frequency per keyword is nearly 2. A look through Table 5 clearly says that two-worded keywords outnumber (~57%) other categories of

Table 3 — Words in keywords

| Words | Frequency | Types of words with respective D(C) | No. of keywords formed |
|---|---|---|---|
| Absorption | 1 | SW (8-C) | 1 |
| Acoustic | 1 | SW (2-C) | 1 |
| Aharonov | 1 | EW | 1 |
| Band | 1 | SW (5-C) | 1 |
| Bohm | 1 | EW | 1 |
| Defect | 1 | SW (4-C) | 1 |
| Effect | 2 | SW (0-C) | 2 |
| Gap | 1 | SW (10-C) | 1 |
| Lattice | 1 | SW (4-C) | 1 |
| Nuclear | 1 | SW (4-C) | 1 |
| Of | 2 | FW | 2 |
| Relaxation | 1 | SW (5-C) | 1 |
| Semiconductor | 1 | SW (3-C) | 1 |
| Spectra | 1 | SW (1-C) | 1 |
| Spin | 1 | SW (10-C) | 1 |
| Surface | 1 | SW (5-C) | 1 |
| Wave | 1 | SW (7-C) | 1 |
| Wide | 1 | SW (4-C) | 1 |

Table 4 — Distribution of concerned articles and keywords for study over the years (2006-2010)

| Year | Vol. No. | No. of articles (A) | No. of distinct keywords (B) | Average no. of distinct keywords per article (B/A) | Total frequency of all distinct keywords (C) | Frequency per keyword (C/B) |
|---|---|---|---|---|---|---|
| 2006 | 32 | 144 | 496 | 3.4 | 1054 | 2.1 |
| 2007 | 33 | 171 | 541 | 3.2 | 1127 | 2.1 |
| 2008 | 34 | 150 | 497 | 3.3 | 903 | 1.8 |
| 2009 | 35 | 149 | 537 | 3.6 | 1059 | 1.9 |
| 2010 | 36 | 155 | 505 | 3.3 | 974 | 1.9 |
| 2006-10 | | 769 | 1155 | 1.5 | 2280 | 2 |

Table 5 — Wordship pattern of keywords over the years (2006-2010)

| Year | Vol. No. | No. of articles | No. of distinct keywords (A) | (Wordship pattern) No. of keywords formed from | | | |
|------|----------|-----------------|------------------------------|-------------|-----------|-------------|----------------------|
| | | | | Single word | Two words | Three words | More than three words |
| 2006 | 32 | 144 | 496 | 107 (22%) | 295 (60%) | 81 (16%) | 13 (3%) |
| 2007 | 33 | 171 | 542 | 120 (22%) | 315 (58%) | 92 (17%) | 14 (3%) |
| 2008 | 34 | 150 | 497 | 119 (24%) | 290 (58%) | 79 (16%) | 9 (2%) |
| 2009 | 35 | 149 | 537 | 138 (26%) | 307 (57%) | 82 (15%) | 10 (2%) |
| 2010 | 36 | 155 | 505 | 125 (25%) | 284 (56%) | 84 (17%) | 11 (2%) |
| 2006-10 | | 769 | 1155 | 276 (24%) | 657 (57%) | 199 (17%) | 23 (2%) |

Table 6 — Statistics of words in keywords over the years (2006-2010)

| Year | Vol. No. | No. of articles | No. of keywords (A) | No. of constituent words (C) | Frequency of words | Average no. of constituent words per keyword (A/C) |
|------|----------|-----------------|---------------------|------------------------------|--------------------|----------------------------------------------------|
| 2006 | 32 | 144 | 496 | 532 | 1002 | 0.93 |
| 2007 | 33 | 171 | 542 | 566 | 1095 | 0.96 |
| 2008 | 34 | 150 | 497 | 534 | 980 | 0.93 |
| 2009 | 35 | 149 | 537 | 546 | 1053 | 0.98 |
| 2010 | 36 | 155 | 505 | 529 | 1003 | 0.95 |
| 2006-10 | | 769 | 1155 | 869 | 2287 | 1.33 |

keywords. The yearwise occurrence and overall relative strength of each category of keywords is presented in Table 5.

The statistics for words in keywords is presented in Table 6. On average, there is nearly one word per keyword has been observed over the years. The overall average over five years (2006-10) reveals presence of 1.3 words per keyword on average.

## Some definitions

### Frequency of words (f)

It is defined as number of times a particular category of keywords occurred and denoted by 'f'. Let us clarify the same taking an example from Table 3, where total number of tetra-contextual (4-C, Ref. Table 1) semantic words are four, e.g. *wide, nuclear, lattice and defect*. Hence the frequency (f) of tetra-contextual semantic words (SW (4-C)) here is 4.

### Number of associations among words (a)

It is defined as total number of associations made by all words belonging to a particular category with other words corresponding to other categories and denoted by 'a'. Suppose in Table 2, the word *wide* forms one association with the word *band*, *nuclear* forms one association with *spin-lattice*, *lattice* forms two associations with *spin* and *relaxation* respectively and finally *defect* forms one association with *absorption-spectra*. Hence, in all five associations have been formed by all tetra-contextual semantic words. Here the value of 'a' is 5.

### Number of keywords formed (k)

It is defined as the total number of keywords formed by the set of words belonging to a particular category. For instance, in Table 2, the tetra-contextual semantic words (SW (4-C)) have been found present in three keywords, viz. *wide band-gap semiconductor*, *nuclear spin-lattice-relaxation effect* and *defect of absorption spectra*. It may thus be stated that the words under category [SW (4-C)] forms three keywords. The value of 'k' here is thus equal to 3.

### Word Association Density (WD(A))

It is defined as the average number of associations developed per unit word belonging to a particular category and denoted by WD(A), which is equal to a/f.

$$WD(A) = a/f \qquad ... (1)$$

### Word Association Coefficient (WC(A))

It is defined as the average number of associations made per unit keyword and denoted by WC(A), which is equal to a/k.

$$WC(A) = a/k \qquad ... (2)$$

### Keyword Formation Density (KD(F))

It is defined as the average number of keywords formed by unit number of word belonging to a particular category and denoted by KD(F), which is equal to k/f.

$$KD(F) = WD(A)/ WC(A) = k/f \qquad ... (3)$$

### Word Association Density Index (WD(A)I)

It is defined as Word Association Density per unit number of keyword and denoted by WD(A)I.

WD(A))I = WD(A)/ k = a/ f*k        ... (4)

**Normalized Word Association Density Index (WD(A))I-N**

This parameter is defined only for semantic words that is defined as Word Association Density Index per unit degree of contextuality or D(C) (Table 1) and denoted by (WD(A))I-N.

(WD(A))I-N = WD(A))I/ D(C) = a/ f*k*D(C)    ... (5)

## Findings

The contextual analysis of words in keywords returns names of different subject domains where the particular words are used. In all, 169 specific subject domains have been observed on analyzing 869 words from two said online reference tools (online dictionary and Wikipedia). These 169 specific domains have been categorized into 24 broad disciplines. These broad disciplines along with specific subject domains and respective frequencies have been presented in Table 7. The frequency of each specific subject is represented by f and the total frequency of a broad discipline is indicated by F = ∑ f. The number of specific subjects in a broad

Table 7 — Broad disciplines and specific subjects in contextual analysis

| Broad disciplines) (alphabetically arranged)and no. of specific subjects (n) contained therein | Specific subjects as obtained from online reference tools with respective frequencies (f) | F = ∑ f | F/n |
|---|---|---|---|
| Agricultural science (3) | Agriculture (12), Apiculture (1), Horticulture (1) | 14 | 4.7 |
| Atmospheric science (1) | Meteorology (16) | 16 | 16 |
| Biological science (18) | Life science (87), Biology (44), Physiology (27), Zoology (25), Botany (22), Genetics (16), Pathology (11), Toxicology (5), Molecular biology (4), Histology (3), Immunology (3), Microbiology (2), Bioinformatics (1), Embryology (1), Entomology (1), Forestry (1), Plant pathology (1), Virology (1) | 255 | 14.2 |
| Chemical science (6) | Chemistry (231),Physical chemistry (18), Organic chemistry (5), Biochemistry (4), Photochemistry (3), Analytical chemistry (1) | 262 | 43.7 |
| Cognitive science (3) | Psychology (34), Philosophy (28), Logic (15) | 77 | 25.7 |
| Computer & information science (4) | Computer science (78), Communication (44), Library & inf. sc (3), Information sc (2) | 127 | 31.8 |
| Earth science (10) | Geology (51), Earth sc (45), Physical geography (25), Mineralogy (12), Geography (7), Hydrology (6), Petrology (6), Oceanography (4), Geodesy (2), Geochemistry (1) | 159 | 16 |
| Engineering science (25) | Engineering (112), Metallurgy (55), Defence sc (37), Mechanical engineering (37), Electrical engineering (26), Mining engineering (24), Printing technology (21), Civil engineering (18), Aerospace engineering (12), Nanotechnology (9),Textile engineering (8), Automotive engineering (7), Aeronautics (6), Naval architecture (5), Chemical engineering (4), Control systems (3), Defence sc (2), Automotive engineering (1), Aviation (1), Chemical technology (1), Design engineering (1), Industrial engineering (1), Petroleum engineering (1), Refrigeration (1), Telecommunications (1) | 394 | 15.8 |
| Environmental science (2) | Ecology (7), Environment (7) | 14 | 7 |
| Home science (1) | Cookery (17) | 17 | 17 |
| Humanities (1) | Literature (15) | 15 | 15 |
| Language (1) | Linguistics (40) | 40 | 40 |
| Management science (4) | Business (13), Accountancy (2), Commerce & business (1), Insurance (1) | 17 | 4.3 |
| Mathematical science (3) | Mathematics (119), Statistics (19), Geometry (5), | 143 | 47.7 |
| Medical science (8) | Medicine (58), Anatomy (27), Dentistry (2), Ophthalmology (2), Pharmacology (2), Gynaecology & obstetrics (1), Psychiatry (1), Surgery (1) | 94 | 11.8 |
| Occultism (1) | Astrology (4) | 4 | 4 |
| Performing and creative arts (19) | Music (50), Sports (38), Fine arts & visual arts (37),Clothing (26), Arts & crafts (19), Architecture (16),Graphic arts (9), Performing arts (11), Photography (9), Theatre (7), Graphics (6), Numismatology (5), Film studies (2), Fishing (2), Cosmetology (1), Fashion designing (1), Hunting (1), Painting (1), Sewing (1) | 242 | 12.8 |

*(Contd.)*

Table 7 — Broad disciplines and specific subjects in contextual analysis (*Contd.*)

| Broad disciplines) (alphabetically arranged)and no. of specific subjects (n) contained therein | Specific subjects as obtained from online reference tools with respective frequencies (f) | F = ∑ f | F/n |
|---|---|---|---|
| Physical science (28) | Physics (359), General physics (188), Electronics (94), Solid state physics (39), Mechanics (36), Atomic physics (24), Optics (19), Quantum mechanics (19), Electromagnetism (17), Nuclear physics (17), Crystallography (16), Fluid mechanics (16), Thermodynamics (10), Acoustics (11), Electricity (10), Cryogenics (5), Geophysics (5), Particle physics (5), Ceramics (3), Spectroscopy (3),Magnetism (2), Photonics (2), Quantum optics (2), Biophysics (1), Heat (1), Nucleonics (1), Plasma physics (1), Quantum chemistry (1) | 904 | 32.3 |
| Religion (2) | Christianity (9), Religion (5) | 14 | 7 |
| Science & technology (in general) (5) | Science (33), Material science (21), Navigation (4),Horology (2), Book binding (1), | 61 | 12.2 |
| Social science (19) | Economics (42), Law (39), Accounting & finance (26), Political sc (17), Social sc (11), Commerce (10), Sociology (10), History (8), Banking & finance (5), Railway transport (5), Anthropology (4), Education (4), Railways transport (2), Culture (1),Journalism & mass communication (1), Library sc & bibliography (1), Road transport (1), Social welfare (1), Transport (2) | 190 | 10 |
| Space science (3) | Astronomy (29), Astrophysics (2), Cosmology (1) | 32 | 10.7 |

discipline is represented by n. The frequency per specific subject is represented by F/n. The relative strengths of the broad disciplines have been presented in Table 8, which shows that the stream of physical science includes 29% words followed by engineering science and chemical science, which include 13% and 9% words respectively. Since the concerned journal belongs to the subject of physics, therefore the topmost position is occupied by physical sciences. The allied subjects in order of proximity of physical sciences are thus engineering science, chemical science, biological science, performing arts, social science, earth science, mathematical science etc. It is interesting to note that the disciplines like performing arts and social science are closer to a facet of physics compared to mathematical science and computer science. The interdisciplinary nature of the facet low temperature physics is thus very prominent from this analysis that is presented in Table 8.

The numerical values of the three fundamental variables for all word categories, i.e. *f, a* and *k* are observed and presented in Table 9. The variation of these three fundamental variables with word categories is shown in Figure 1. It is observed that all three variables possessed highest values for 2-C category semantic words and lowest values for 14-C category semantic words. The values are also fairly large for eponymous words that indicates leading role of the same in keyword formation. The eponymous words are derived from the names of the scientists.

Table 8 — Ranking of broad disciplines by total frequency F

| Rank | Broad disciplines | F = ∑ f | Percentage |
|---|---|---|---|
| 1 | Physical science | 904 | 29.25 |
| 2 | Engineering science | 394 | 12.75 |
| 3 | Chemical science | 262 | 8.48 |
| 4 | Biological science | 255 | 8.25 |
| 5 | Performing arts | 240 | 7.76 |
| 6 | Social science | 190 | 6.15 |
| 7 | Earth science | 159 | 5.14 |
| 8 | Mathematical science | 143 | 4.63 |
| 9 | Computer & information science | 127 | 4.11 |
| 10 | Medical science | 94 | 3.04 |
| 11 | Cognitive science | 77 | 2.49 |
| 12 | Science & technology | 61 | 1.97 |
| 13 | Language | 40 | 1.29 |
| 14 | Space science | 32 | 1.04 |
| 15 | Home science | 17 | 0.55 |
| 15 | Management science | 17 | 0.55 |
| 16 | Atmospheric science | 16 | 0.52 |
| 17 | Agricultural science | 14 | 0.45 |
| 17 | Environmental science | 14 | 0.45 |
| 17 | Humanities | 14 | 0.45 |
| 17 | Religion | 14 | 0.45 |
| 18 | Occultism | 4 | 0.13 |
| 19 | Creative arts | 2 | 0.06 |

The numerical values of the five parameters defined in Equations (1) to (5) for all word categories have been calculated from these three variables, i.e. 'f', 'a' and 'k'. The values of the five parameters have been presented in Table 9.

Table 9 — Values of some word association parameters for different word categories

| | D(C) | f | a | k | WD(A) | WC(A) | KD(F) | WD(A)I | WD(A)I-N |
|---|---|---|---|---|---|---|---|---|---|
| Semantic Words (SW) | 0-C | 38 | 100 | 99 | 2.63 | 1.01 | 2.61 | 0.027 | |
| | 1-C | 133 | 200 | 166 | 1.52 | 1.20 | 1.25 | 0.009 | 0.009 |
| | 2-C | 164 | 335 | 285 | 2.06 | 1.18 | 1.74 | 0.007 | 0.004 |
| | 3-C | 125 | 235 | 180 | 1.90 | 1.31 | 1.44 | 0.011 | 0.004 |
| | 4-C | 72 | 227 | 210 | 3.20 | 1.08 | 2.92 | 0.015 | 0.004 |
| | 5-C | 48 | 108 | 107 | 2.30 | 1.01 | 2.23 | 0.021 | 0.004 |
| | 6-C | 46 | 156 | 150 | 3.39 | 1.04 | 3.26 | 0.023 | 0.004 |
| | 7-C | 31 | 115 | 114 | 3.71 | 1.01 | 3.68 | 0.033 | 0.005 |
| | 8-C | 42 | 101 | 109 | 2.41 | 0.93 | 2.60 | 0.022 | 0.003 |
| | 9-C | 29 | 96 | 91 | 3.31 | 1.05 | 3.14 | 0.036 | 0.004 |
| | 10-C | 19 | 61 | 57 | 3.21 | 1.07 | 3.00 | 0.056 | 0.006 |
| | 11-C | 7 | 43 | 43 | 6.14 | 1.00 | 6.14 | 0.143 | 0.013 |
| | 12-C | 7 | 85 | 85 | 12.14 | 1.00 | 12.14 | 0.143 | 0.012 |
| | 13-C | 8 | 25 | 25 | 3.13 | 1.00 | 3.13 | 0.125 | 0.010 |
| | 14-C | 4 | 6 | 6 | 1.50 | 1.00 | 1.50 | 0.250 | 0.018 |
| | 17-C | 4 | 8 | 8 | 2.33 | 1.00 | 2.00 | 0.292 | 0.017 |
| | 19-C | 4 | 16 | 16 | 5.00 | 1.00 | 4.00 | 0.313 | 0.016 |
| | AC | 10 | 9 | 9 | 0.90 | 1.00 | 0.90 | 0.100 | |
| | EW | 73 | 86 | 71 | 1.18 | 1.21 | 0.97 | 0.017 | |
| | FW | 7 | 8 | 8 | 1.14 | 1.00 | 1.14 | 0.143 | |

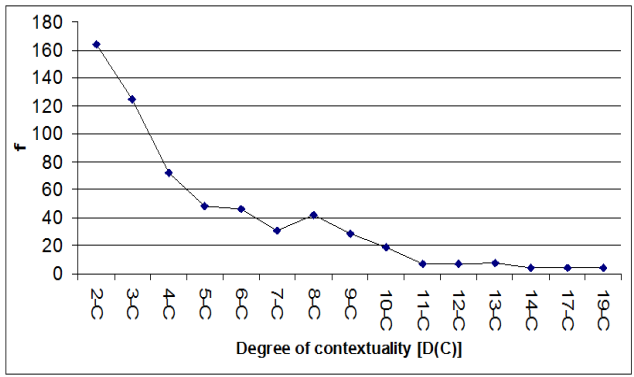

Fig. 1 — Variation of three fundamental variables (f, a & k) with word categories

It is observed that the Word Association Coefficient (WC(A)) remains almost constant for all word categories as shown in the sixth column of Table 9. The value of (WC(A)) is nearly one here. Since WC(A) = a/k, therefore it may be inferred that 'a' (no. of associations) is directly proportional to 'k' (no. of keywords) here. The values of 'a' and 'k' are in the same order, i.e. nearly equal as a/k ~ 1. Hence an increasing tendency of associations among the words enhances the number of keywords also. The Word Association Coefficient (WD(A)) varies for different word categories (Table 9) ranging from 12.14 to 0.9. The highest value is observed for 12-C semantic words and the lowest one is observed for the acronyms. Almost similar type of variation pattern has been observed for Keyword Formation Density (KD(F)) also.

The Word Association Density Index (WD(A))I varies for different word categories (Table 9) ranging from 0.313 to 0.007. The highest value is observed for 19-C semantic words and the lowest one is observed for the 2-C semantic words. This value is same for form words, 11-C and 12-C semantic words. The Normalized Word Association Density Index (WD(A))I-N varies for different word categories (Table 9) ranging from 0.018 to 0.003. The highest value is observed for 14-C semantic words and the lowest one is observed for the 8-C semantic words. This value is almost identical for semantic words of 2-C to 9-C categories, which ranges from 0.003 to 0.005. Of these, the semantic words of 2-C to 6-C and 9-C categories have exactly same values, i.e. 0.004. The range of variation of normalized (WD(A))I is much less compared to the same for (WD(A))I. Hence the parameter (WD(A))I-N may be considered as nearly constant for all word categories.

Now, (WD(A))I-N = WD(A))I/ D(C) = a/ f*k*D(C) (Equation (5)) and it has been found that (a/k) remains almost constant for all categories. Since the values of (WD(A))I-N also remains more or less constant therefore [a/ {f*k*D(C)}] would be a constant quantity. Hence, [1/ {f* D(C)}] will also be nearly constant, or it may be inferred that 'f' is inversely proportional to D(C). The frequency of words will decrease with the increase of degree of contextuality. This trend is also clear from Figure 2 except 0-C, 1-C and 8-C word categories. The words that are considered relevant in larger number of subjects comparatively less appeared in the keywords.

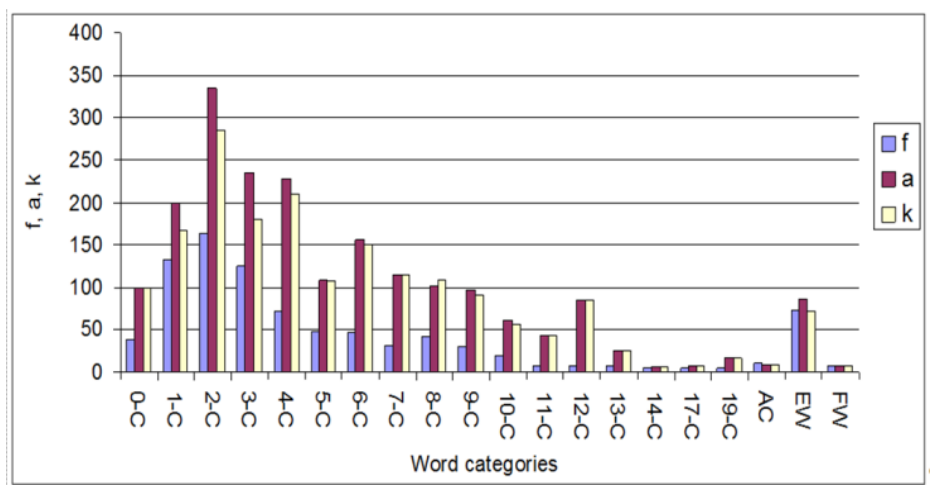

Fig. 2 — Variation of Degree of Contextuality [D(C)] with frequency of words (f)

## Conclusion

The journal used for this study belongs to *physics*, but the constituent words in assigned keywords have been found relevant in different subjects other than physics. In particular, fairly large number of words occurred from the broad disciplines like biological science, performing arts and social science, which is very interesting feature. The wordship pattern analysis shows the dominance of two-worded keywords. The words with lower degree of contextuality participate in formation of major number of keywords compared to words with higher degrees of contextualities. A considerably large number of words with lower degree of contextuality are coupled with eponymous words to form important subject keywords.

The eponymous words are thus very dynamic to form keywords in *low temperature physics*. It is observed that number of word associations is directly proportional to number of keywords, and frequency of words is inversely proportional to degree of contextuality, i.e. more number of associations among words will tend to formation of larger number of keywords and words with higher degree of contextuality are less compared to the same of lower degree of contextuality. The increase in number of variety keywords with number of associations indicates increasing emergence of new concepts in this area. The semantic words of 2-C category are highest in number compared to all other categories. The words relevant in larger number of subjects rarely form domain-specific keywords, whereas the words generally used in the context of few number of subjects (two, three or four) are very prolific in forming domain-specific keywords. This study has presented a model for word analysis of keywords in any subject area to understand the nature of constituent words. The allied subject areas of a specific subject domain are understood from the context pattern of words in keywords. This study may be extended to other subject domains.